\documentclass[%
superscriptaddress,
twocolumn,
amsmath,amssymb,
aps,
pra,
floatfix,
]{revtex4-1}

\usepackage{graphicx}
\usepackage{dcolumn}
\usepackage{bm}
\usepackage{physics}
\usepackage{mathrsfs}
\usepackage{comment}
\usepackage{amsmath}
\usepackage[T1]{fontenc}

\begin{document}


\title{Resonator-assisted single molecule quantum state detection}
\author{Ming Zhu}
\affiliation{
 Department of Physics and Astronomy, Purdue University, West Lafayette, IN 47907, USA
}
\author{Yan-Cheng Wei}
\affiliation{
 Department of Physics and Astronomy, Purdue University, West Lafayette, IN 47907, USA
}
\affiliation{
 Department of Physics, National Taiwan University, Taipei 10617, Taiwan
}
\author{Chen-Lung Hung}
\email{Email: clhung@purdue.edu}
\affiliation{
 Department of Physics and Astronomy, Purdue University, West Lafayette, IN 47907, USA
}
\affiliation{
 Purdue Quantum Science and Engineering Institute, Purdue University, West Lafayette, IN 47907, USA
}

\begin{abstract}
We propose a state-sensitive scheme to optically detect a single molecule without a closed transition, through strong coupling to a high-Q whispering-gallery mode high-Q resonator. A background-free signal can be obtained by detecting a molecule-induced transparency in a photon bus waveguide that is critically coupled to the resonator, with a suppressed depumping rate to other molecular states by the cooperativity parameter $C$. We numerically calculate the dynamics of the molecule-resonator coupled system using Lindblad master equations, and develop analytical solutions through the evolution of quasi-steady states in the weak-driving regime. Using Rb$_2$ triplet ground state molecules as an example, we show that high fidelity state readout can be achieved using realistic resonator parameters. We further discuss the case of multiple molecules collectively coupled to a resonator, demonstrating near-unity detection fidelity and negligible population loss.
\end{abstract}
\date{\today}
\maketitle

\section{Introduction}
The ability to determine the state of a single quantum emitter is essential for quantum information processing. Resonance fluorescence imaging is a convenient and powerful method. During imaging, exciting closed optical transitions ensures that a quantum emitter scatters a large number of photons without leaving a specific ground state, thus making it possible to achieve state-sensitive detection with a high signal-to-noise ratio, even with low photon collection efficiency. This approach has been widely employed in various quantum systems, such as cold atoms \cite{bochmann2010lossless, gehr2010cavity, fuhrmanek2011free}, trapped ions \cite{olmschenk2007manipulation, myerson2008high}, nitrogen vacancy centers \cite{neumann2010single, robledo2011high} and quantum dots \cite{vamivakas2010observation}. 

Many promising quantum systems are not suited for fluorescent imaging due to their energy structure. For instance, cold molecules have a wide range of applications in quantum chemistry \cite{bohn2017cold} and quantum computation \cite{demille2002quantum}. However, most of the molecules have no real optical cycling transitions due to a myriad of rovibrational levels accessible in a radiative decay process. For specific kinds of molecules, one may find a transition with near-unity Franck-Condon factor for a target ground state, and use multiple lasers to drive an approximately closed transition in a manageable collection of ground and excited states. For instance, SrF \cite{shuman2010laser} or CaF \cite{anderegg2017radio} molecules can be laser cooled and trapped in a magneto-optical trap. 

In general, alternative state detection method is needed for molecules without optically closed transitions. Resonance-enhanced multiphoton ionization is adopted for molecule detection \cite{gabbanini2000cold}, which ionizes the molecules and detects the subsequent ions. This method is nevertheless completely destructive. Direct absorption imaging of molecules is also reported \cite{wang2010direct}, when a molecular ensemble has a high optical density.\par

Cavity quantum electrodynamics opens a new way to implement state detection of a single molecule. Cavity-controlled light-matter interaction enables the manipulation of molecular photon emission properties.  When a molecule is located inside a cavity or near a resonator which is tuned to the molecular resonance of interest, the branching ratio of decay into irrelevant states may be greatly suppressed \cite{lev2008prospects}, allowing the interaction with resonator photons for an extended period of time. In addition, considering the molecules directly emit photons into the resonator mode(s), the signal photon collection efficiency may be significantly enhanced compared to the case of emission in freespace. \par

\begin{figure}[t]
    \centering
    \includegraphics[width=1 \columnwidth]{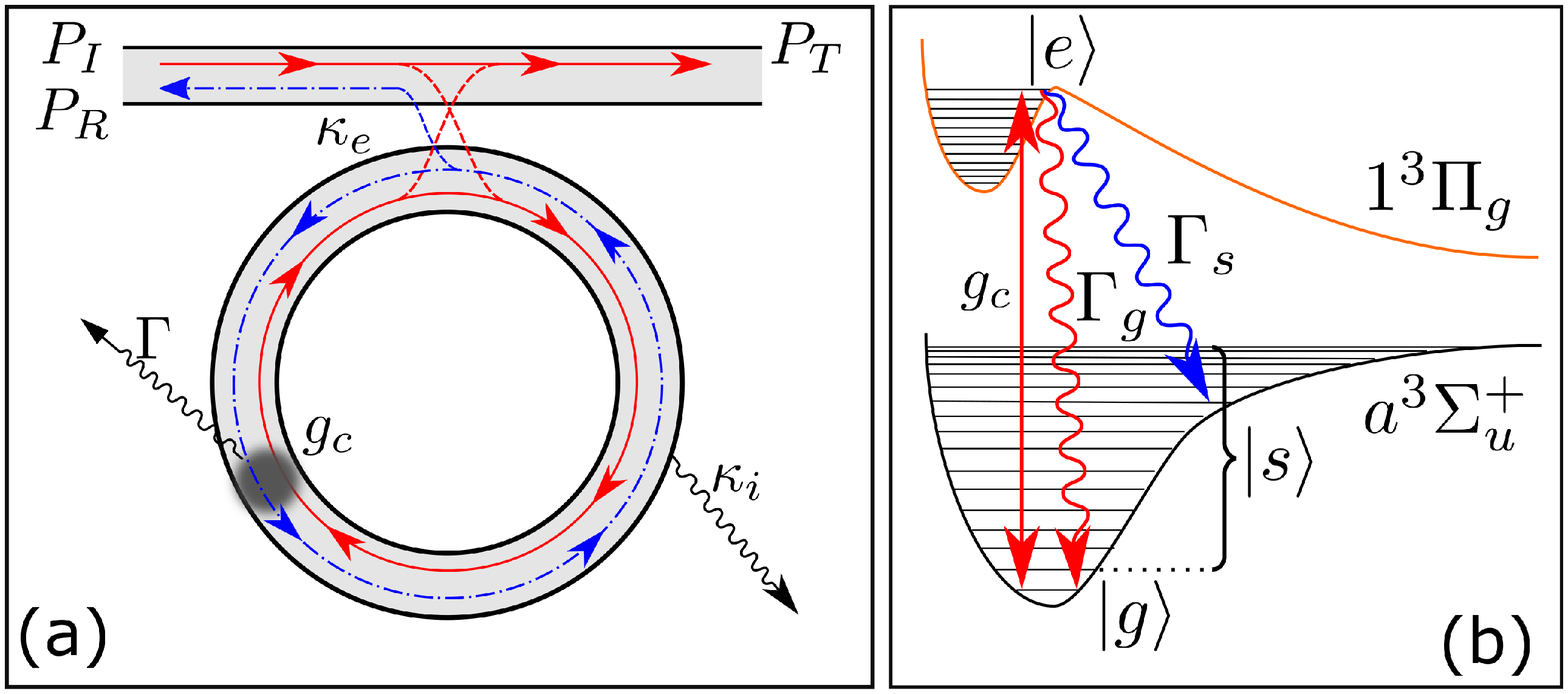}
    \caption{(color online) Schematic illustration of the investigated system. (a) Optical setup marked by basic rates. A resonator supports both clockwise (red solid arrows) and counter-clockwise (blue dashed arrows) circulating whispering-gallery modes (WGMs), and is excited by a bus waveguide. (b) Sample transition in Rb$_{2}$ molecule between the rovibrational ground state $\ket{g}$ in the $a^{3}\Sigma_{u}^{+}$ potential and a molecular excited state $\ket{e}$ in the $1^{3}\Pi_{g}$ potential. State $\ket{s}$ represents a collection of all other rovibrational levels in the $a^{3}\Sigma_{u}^{+}$ potential. A molecule in state $\ket{g}$ (shaded circle in (a)) interacts with the WGM(s) at a coupling rate $g_c$.} 
    \label{fig: ring resonator}
\end{figure}

In this paper, we consider a scheme to detect the quantum state of molecules without a closed transition by utilizing high-finesse Fabry-Perot cavities \cite{hood2001characterization} or high-Q whispering-gallery mode (WGM) resonators \cite{vernooy1998high, spillane2005ultrahigh, pollinger2009ultrahigh, tien2011ultra, chang2019microring}. Our scheme is inspired by the pioneering experiments for detecting single atoms falling through a microtoroidal resonator \cite{aoki2006observation, shomroni2014all}, and probing trapped single atoms inside a mirror cavity \cite{boozer2006cooling, khudaverdyan2009quantum}, a fiber-based cavity \cite{gehr2010cavity, kato2015strong}, as well as in the vicinity of a photonic crystal cavity \cite{thompson2013coupling, goban2014atom}. In particular, we consider the transmission of a bus waveguide critically coupled to WGMs in a micro-ring resonator [Fig.~\ref{fig: ring resonator}(a)], displaying molecule-induced transparency on resonance due to strong light-molecule interaction. The proposed detection scheme is background-free, as there is no waveguide transmission unless a molecule in the target state couples to the resonator. The maximum scattered photon number is enhanced by the cooperativity parameter $C$ before the molecule is completely pumped away from its initial state. We numerically simulate this open quantum system with a Lindblad master equation and derive analytical solutions in the weak-driving regime. We then extend the single-molecule model to a multi-molecule case considering collective effect. Similar scheme can be realized in monitoring the off-resonant transmissivity in the case of a Febry-Perot cavity (Appendix~\ref{app: FB cavity}). Our scheme can also be applied to single-shot state readout for other quantum emitters \cite{sun2016single}. 

For illustration purposes, we take Rb$_{2}$ molecule as an example. The relevant internuclear potentials are plotted in Fig.~\ref{fig: ring resonator}(b), in which a single-photon, short-range photoassociation has been utilized for ground state molecule synthesis directly from cold atoms \cite{bellos2011formation}. Using an optical cavity or a photonic crystal waveguide to radiatively enhance the synthesis efficiency has recently been discussed \cite{perez2017ultracold, kampschulte2018cavity}. These recent developments makes Rb$_{2}$ a good candidate to demonstrate resonator-enhanced single molecule detection.

The paper is organized as follows. Section~\ref{sec:setup} reviews the optical setup of the resonator as well as a simplified energy structure of the Rb$_2$ molecule. In Sec.~\ref{sec:one mode}, we derive a formalism to calculate single molecule dynamics with one resonator mode. In Sec.~\ref{sec:two mode}, the discussion is extended to the coupling between a single molecule and two resonator modes. In Sec.~\ref{sec:multi}, we discuss the multi-molecule dynamics by taking into account collective effects.

\section{\label{sec:setup} The Optical Setup and Molecular Model}
We first introduce the optical system, a bus waveguide coupled to an empty micro-resonator as shown in Fig.~\ref{fig: ring resonator}(a), which supports WGMs that circulate either in the clockwise (CW) or the counter-clockwise (CCW) orientation along the resonator. We assume an input field of power $P_I$ is injected from one end of the bus waveguide, and analyze the transmission power $P_T$ and reflection power $P_R$. In the following cases, we assume that back-scattering in the resonator, which couples the CW and CCW modes, occurs at a rate much smaller than the total resonator loss rate $\kappa$, and thus there is negligible mode mixing. Here, $\kappa=\kappa_e+\kappa_i$, where $\kappa_i$ is the intrinsic photon loss rate and $\kappa_e$ is the waveguide-resonator coupling rate. Due to the phase matching condition, when the resonator couples light from single end of the bus waveguide shown in Fig.~\ref{fig: ring resonator}(a), only one mode (CW WGM,  illustrated as solid line) can be excited and the resonator can be treated like a single mode cavity. Reflection stays at zero ($P_R=0$) due to the absence of CCW WGM excitation.

\begin{figure}[t]
    \centering
    \includegraphics[width=0.8\columnwidth]{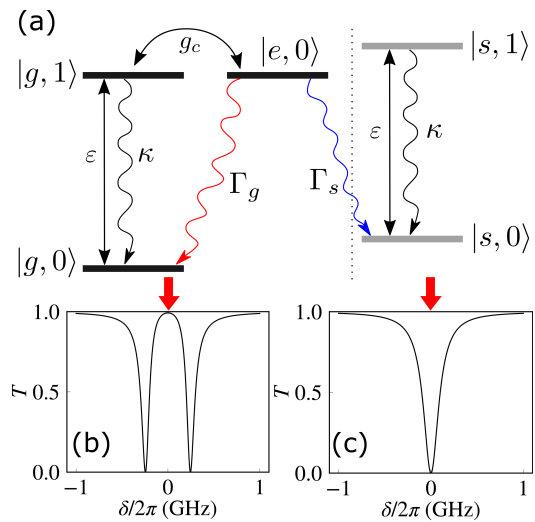}
    \caption{(a) Simplified energy level structure for one molecule coupled to single resonator mode in the weak-driving regime for $\omega_{m} = \omega_{c}$. (b) Transmission spectrum of a molecule-coupled resonator with $\Gamma_s=0$. (c) Transmission spectrum of an empty resonator. Background-free measurement can be performed on resonance $\delta\equiv\Delta_{\textrm{cl}} =0$.}
    \label{fig: one_weak driving}
\end{figure}

To achieve background-free molecule detection, we consider a critically coupled resonator ($\kappa_i=\kappa_e$) for zero waveguide transmissivity on resonance ($\Delta_{\text{cl}} = 0$). The bus waveguide transmissivity is evaluated by solving the standard Heisenburg-Langevin equation with the single-mode Hamiltonian of an empty resonator
\begin{equation}
\label{eq: H_ring}
\hat{H}_{0} =\Delta _{\text{cl}} \hat{a}^{\dagger } \hat{a}+i\left( \varepsilon \hat{a}^{\dagger } -\varepsilon ^{*} \hat{a}\right),
\end{equation}
where $\Delta_{\text{cl}} = \omega_{c}-\omega_{l}$ is the detuning between the resonant mode frequency $\omega_{c}$ and the external driving frequency $\omega_{l}$, $\hat{a}^{(\dag)}$ represents the annihilation (creation) operator of the (CW) resonator mode, excited from the bus waveguide at a rate coefficient $\varepsilon=i \sqrt{2 \kappa_{e} \mathcal{I}}$ and $\mathcal{I} = P_{I}/\left( \hbar \omega_{l} \right)$ is the waveguide photon input rate. At steady state, the expectation value of intra-resonator field is found to be $\expval{\hat{a}}=\varepsilon/\left( i \Delta _{\textrm{{cl}}} +\kappa \right)$.
The bus waveguide transmission is the interference between the input field $\sqrt{\mathcal{I}}$ and the out-coupled field from the resonator $i \sqrt{2 \kappa_{e}} \expval{a}$, resulting in a transmissivity \cite{haus1984waves}
\begin{equation}  T = \frac{P_{T}}{P_{I}} =\left| 1+i\sqrt{\frac{2\kappa _{e}}{\mathcal{I}}} \langle \hat{a}\rangle \right| ^{2}.
\label{eq: transmission}
\end{equation}
As shown in Fig.~\ref{fig: one_weak driving}(c), at critical coupling the waveguide transmissivity, $T=\abs{\Delta_{\text{cl}}/(\kappa + i \Delta_{\text{cl}})}^2$, drops to zero on resonance. As we discuss in the following, the waveguide transmissivity will be greatly modified when a molecule is present and couples to the resonator. This establishes our scheme to realize background-free molecule detection.

To model the radiative dynamics of a resonator-coupled molecule, we treat it as an effective three-level system, as illustrated in Fig.~\ref{fig: ring resonator}(b). While radiative decay processes can couple a molecular excited state $\ket{e}$ to a collection of (electronic) ground states of different rovibrational energy levels, the molecule-resonator coupling can in principle involve only one designated rovibrational state $\ket{g}$ when the resonator frequency $\omega_c$ is aligned with the transition frequency $\omega_{m}$ and the resonator linewidth ($\lesssim O(1)~$GHz) is smaller than the relevant rovibrational energy level spacing ($\gtrsim O(10)~$GHz) by over an order of magnitude. A third state $\ket{s}$ denotes all other uncoupled states that can accumulate population from spontaneous decay. Given a Franck-Condon factor $f_\mathrm{FC}$ between states $\ket{e}$ and $\ket{g}$, the spontaneous decay rate to $\ket{g}$ is $\Gamma_g=f_\mathrm{FC}\Gamma$ and the decay rate to $\ket{s}$ is $\Gamma_s = (1-f_\mathrm{FC})\Gamma$, where $\Gamma$ is the total decay rate of the excited state $\ket{e}$.

We denote the coherent coupling rate between the molecule and the resonator mode as $\displaystyle g_c = \sqrt{3 \Gamma_{g} c^3/2 V\omega_{m}^{2}}$, where $c$ is the speed of light, $\displaystyle V=\int \epsilon(\mathbf{r})|\mathcal{E}(\mathbf{r})|^2d\mathbf{r}/ \left|\mathcal{E}\left(\mathbf{r}_\mathrm{mol}\right)\right|^2$ is the mode volume, $\epsilon(\mathbf{r})$ is the dielectric function, $\mathcal{E}\left(\mathbf{r}\right)$ is the mode field strength, and $\mathbf{r}_\mathrm{mol}$ is the molecular position. $g_c$ is position dependent due to the mode field intensity variation near the resonator dielectric surface. We assume that molecules are trapped in close proximity of a resonator and hence $g_c$ is a constant. 

\begin{table}[b]
\caption{\label{tab: parameters} Sample parameters adopted for the investigated platform, using Rb$_2$ ($\ket{g}$: $\nu'=0$ in $a^3\Sigma^+_u$, $\ket{e}$: $\nu=8$ in $1^3\Pi_g^+$)\cite{bellos2011formation}}
\begin{ruledtabular}
\begin{tabular}{lcc}
\textrm{Parameter}&
\textrm{Symbol}&
\textrm{Value}\\
\colrule
Total spontaneous decay rate ($\ket{e}$) & $\Gamma$      & $2 \pi \times 12\mathrm{MHz}$ \\
Franck-Condon factor ($\ket{e}\Longleftrightarrow\ket{g}$)         & $f_{FC}$      & 0.37 \\
\colrule
Photon input rate            & $\mathcal{I}$ & $1 \mathrm{MHz}$\\
Resonator intrinsic loss rate          & $\kappa_{i}$  & $2 \pi \times 50\mathrm{MHz}$ \\
External coupling rate       & $\kappa_{e}$  & $2 \pi \times 50\mathrm{MHz}$ \\
Sample cooperativity parameter      & $C$           & 50 \\
Resonator coupling rate ($C=50$) & $g_c$     & $2 \pi \times 245\mathrm{MHz}$ \\
\end{tabular}
\end{ruledtabular}
\end{table}

Table \ref{tab: parameters} lists typical resonator and molecule parameters used in the numerical and analytical calculations throughout this paper. We take Rb$_{2}$ molecules coupled to a micro-ring resonator as an example. The relevant internuclear potentials are plotted in Fig.~\ref{fig: ring resonator}(b), where $\ket{g}$ stands for the rovibrational ground state of interest in the $a^3\Sigma^+_u$ triplet potential, and $\ket{e}$ represents an excited state in the $1^3\Pi_g$ potential (vibrational level $\nu = 8$, and angular momentum quantum number $J = 1$) \cite{bellos2011formation}. Due to the finite Frank-Condon overlap between $\ket{g}$ and $\ket{e}$, this transition has been utilized for ground state molecule synthesis directly from cold atoms via a single-photon short-range photoassociation to state $\ket{e}$, followed by spontaneous decay into $\ket{g}$ \cite{bellos2011formation}. On the other hand, our model optical system is adapted from a recent report on high-Q micro-ring resonators \cite{chang2019microring}, where we assume that high quality factor $Q>10^{5}$ and large single-photon vacuum Rabi frequency $2g_c \sim 2\pi \times 500~$MHz can be simultaneously realized to achieve a large coorperativity parameter $C \equiv g_c^2/\kappa\Gamma$, which is the key parameter to achieve high single-molecule detection sensitivity.

\section{\label{sec:one mode}Single Molecule Dynamics coupled with one resonator mode}

We now analyze the recovery of bus waveguide transmissivity on resonance with the presence of a single molecule as shown in Fig.~\ref{fig: one_weak driving}(b). We begin with the first scenario where a molecule couples to a single resonator mode. This applies to the case when a molecule, spin-polarized in a stretched state, couples only to a circularly polarized WGM, and cannot emit photons into the other WGM because of its opposite circular polarization state \footnote{The WGMs are circularly polarized since they are traveling wave modes with strong evanescent field outside the waveguide. A $\pm\pi/2$ degree out-of-phase axial field component, dictated by the transversality of the Maxwell's equation and time reversal symmetry, gives the CW and CCW WGMs opposite circular polarization states.}. The single-mode light-molecule interaction Hamiltonian $\hat{H}_{1}$ can be written as
\begin{equation}
    \hat{H}_{1}= \Delta_{\text{ml}} \hat{\sigma}_{+}\hat{\sigma}_{-} + i g_c \left( \hat{a}^{\dag} \hat{\sigma}_{-} - \hat{a} \hat{\sigma}_{+} \right),
\label{eq: H1}
\end{equation} 
where $\Delta_{\text{ml}}=\omega_{m}-\omega_{l}$, $\hat{\sigma}_{-} = \ket{g}\bra{e}$, and $\hat{\sigma}_{+} = \ket{e}\bra{g}$. 

Taking into account of the resonator loss and the molecule spontaneous emission, the master equation of the full system is written as
\begin{equation}
    \begin{aligned}
    \frac{d\rho}{dt} =& -i \left[ \hat{H_{0}}+\hat{H_{1}}, \rho \right] + \\ &2\kappa\mathcal{L}[\hat{a}]\rho +\Gamma_{g} \mathcal{L}[\hat{\sigma}_{-}]\rho + \Gamma_{s} \mathcal{L}[\hat{\sigma}_{-}^{\prime}]\rho,
\end{aligned}
\label{eq: one_master equation}
\end{equation}
where $\rho$ is the density matrix of the molecule and photon system and the Lindblad operators take the form of $\mathcal{L}[\hat{b}] \rho=\hat{b} \rho \hat{b}^{\dagger}-\frac{1}{2} \hat{b}^{\dagger} \hat{b} \rho-\frac{1}{2} \rho \hat{b}^{\dagger} \hat{b}$ and $\hat{\sigma}_{-}^{\prime} = \ket{s}\bra{e}$. As shown in Table~\ref{tab: parameters}, both the coherent coupling rate $g_c$ and the total resonator loss rate $\kappa = \kappa_i + \kappa_e = 2 \kappa_e$ are at least an order of magnitude larger than the molecule spontaneous decay rates $\Gamma_{g(s)}$, thus allowing us to utilize fast resonator-molecule dynamics for state detection before losing the population into the uncoupled states $\ket{s}$. 

In the limit of single excitation, the resonator and the molecule form an effective five-level system with states denoted by $\ket{m,n}$ as shown in Fig.~\ref{fig: one_weak driving}(a), where $m$ represents the molecular quantum state and $n=0, 1$ is the photon number in the WGM. 
Three coupled states $\ket{e,0}$ and $\ket{g,0(1)}$ on the left of Fig.~\ref{fig: one_weak driving}(a) evolve under the cavity QED Hamiltonian $\hat{H_0}+\hat{H_1}$ that can quickly establish a quasi-steady state under external driving. Population in this subsystem then slowly decays into the uncoupled states illustrated in the right part of Fig.~\ref{fig: one_weak driving}(a), evolving as an empty resonator.

\begin{figure}[t]
    \centering
    \includegraphics[width=1 \columnwidth]{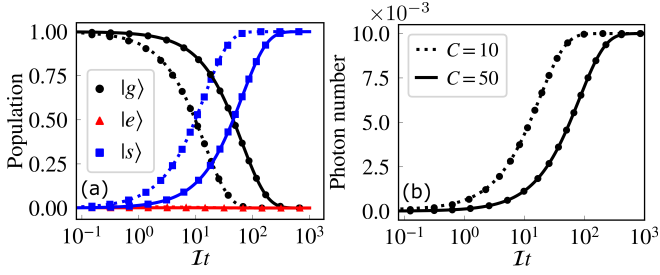}
    \caption{(color online) Time evolution of (a) state populations and (b) resonator photon number $\expval{a^{\dag}a}$ under a resonant weak-driving ($\delta$=0), evaluated analytically (lines) and numerically (symbols) using cooperativity parameters $C = 10$ (dotted lines), and 50 (solid lines), respectively. For other system parameters used in this and the remaining figures, see Table~\ref{tab: parameters}.}
    \label{fig: single dynamics}
\end{figure}

Figure~\ref{fig: one_weak driving}(b) shows the steady-state transmissivity in an ideal cavity QED system for $\Gamma_s=0$. The transmission spectrum is evaluated by substituting the expectation value of the field amplitude $\langle \hat{a} \rangle$ in Eq.~(\ref{eq: transmission}) with the steady-state solution
\begin{equation}
    \expval{\hat{a}}=\frac{i\Delta_{\text{ml}}+\Gamma/2}{g_c^{2}+(i \Delta_{\text{cl}} +\kappa)(i \Delta_{\text{ml}} + \Gamma/2)}\varepsilon.
\label{eq: one_steady_a}
\end{equation}
Recalling that the WGM frequency $\omega_{c}$ is aligned to the molecular transition frequency $\omega_{m}$, we define detuning $\delta \equiv \Delta_{\text{cl}}  \equiv \Delta_{\text{ml}}$. When the cooperativity $C \gg 1$, the transmissivity at zero detuning ($\delta=0$),
\begin{equation}
    T_0=\left|1-\frac{\kappa \Gamma/2}{g_c^2+\kappa\Gamma/2}\right|^2 \approx 1-\frac{1}{C},
\end{equation}
nearly recovers to unity. This effect can be understood as the interference of two molecule-photon dressed states that results in a molecule-induced transparency window, similar to an electromagnetically-induced transparency (EIT). The EIT-like effect contrasts the vanishing transmissivity of an empty resonator critically coupled to the bus waveguide as shown in Fig.~\ref{fig: one_weak driving}(c). This forms a highly sensitive scheme for quantum state detection similarly found in \cite{aoki2006observation}.

In the realistic case of $\Gamma_s\neq 0$, transmission can only recover for a finite period of time. One expects that the transmission spectrum evolves transiently from an EIT-like curve of Fig.~\ref{fig: one_weak driving}(b) to the empty resonator case as shown in Fig.~\ref{fig: one_weak driving}(c). Nevertheless, strong resonator coupling suppresses the excited state $\ket{e}$ population, resulting in a much reduced decay rate into the uncoupled states $\ket{s}$ compared to the free-space decay rate $\Gamma_{s}$. Finite transmission through the bus waveguide can thus be collected for a finite time period for molecular state detection. 

Using the separation of time scales, we find the analytical solution for the quasi-steady density matrix $\rho^{ss}$ of the system, as detailed in Appendix~\ref{app: single}, and evaluate the population transfer rate $D$ to the empty resonator. In the weak-driving regime ($\displaystyle \abs{\epsilon} \ll \kappa, g_c^2/\kappa$), we find the population of the molecule-resonator dressed state primarily resides in $\ket{g,0}$. The transfer rate $D$ is greatly suppressed due to a small population in $\ket{e,0}$. Based on Eq.~(\ref{eq: e0_pop}), We find
\begin{equation}
D(\delta) =  \frac{\rho_{e0,e0}^{ss}}{\rho_{g0,g0}^{ss}} \Gamma_{s} = \frac{ g_c^2 \kappa \mathcal{I}}{\abs{ g_c^2 +(i \delta +\kappa )\left(i \delta +\frac{\Gamma }{2}\right)} ^{2}}\Gamma_{s}, \label{eq: decay rate_detuning}
\end{equation}
where $\rho_{e0,e0}~(\rho_{g0,g0})$ is the population in the excited state $\ket{e,0}$ (ground state $\ket{g,0}$), and the superscript $ss$ stands for steady state. $D/\mathcal{I}$ also represents the probability for the dressed state to decay into $\ket{s,0}$. 

At zero detuning, we find the decay rate
\begin{equation}
    D_\mathrm{res}=\frac{C}{(1/2+C)^2}\frac{\Gamma_s}{\Gamma}\mathcal{I}. 
\label{eq: decay rate}
\end{equation} 
Comparing $D_\mathrm{res}/\mathcal{I}$ with the depumping probability in freespace $\Gamma_s/\Gamma=(1-f_{FC})$, the resonator-enhanced $\ket{e}$-$\ket{g}$ transition enjoys a large suppression factor $\sim 1/C$ for depumping into the uncoupled states when $C \gg 1$. As we will show in Eq.~(\ref{eq: one_max_number}), this factor suggests that around $\sim C/(1-f_{FC})$ photons may be transmitted through the bus waveguide before the system is converted into an empty resonator and again blocks all resonant input photons. 

We validated the slow-transfer model with full numerical calculations \cite{Qutip} using the master equation Eq.~(\ref{eq: one_master equation}) and the parameters listed in Table~\ref{tab: parameters};  see Fig.~\ref{fig: single dynamics} for an example. The numerical result shows negligible differences from the analytical model in the mean resonator photon number $\langle \hat{a}^\dag \hat{a}\rangle$ [Eq.~(\ref{eq: photon_number_one_mode})], as well as the state populations
\begin{equation}
\begin{aligned}
    P_{g}(t) &\approx \rho_{g0,g0}(t)=\exp(-D_\mathrm{res}t), \\
    P_{s}(t) &\approx \rho_{s0,s0}(t)=1-\exp(-D_\mathrm{res}t),\\
    P_{e}(t) &= P_g(t) D_\mathrm{res}/\Gamma_s.
\end{aligned}
\label{eq: population}
\end{equation}
In the following discussions, we will mainly present  our analytical analysis.

\begin{figure}[t]
    \centering
    \includegraphics[width=0.8 \columnwidth]{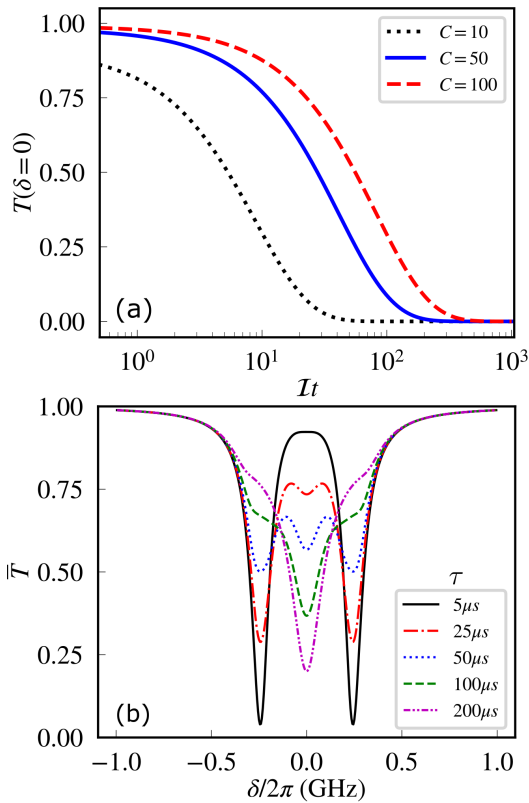}
    \caption{(a) Time evolution of resonant transmissivity $T(0)$. 
    (b) Time-averaged transmission spectra $\overline{T}(\delta,\tau)$ evaluated under various time intervals (0, $\tau$) with $C=50$.} 
    \label{fig: transmission-one mode}
\end{figure}

Now we derive the transmission spectrum by evaluating the time evolution of the quasi-steady resonator field $\expval{\hat{a}} = \rho^{ss}_{g1,g0} + \rho^{ss}_{s1,s0}$ as detailed in Eq.~(\ref{eq: exp_a_one_mode}), where $\rho_{g1,g0}$ and $\rho_{s1,s0}$ are the off-diagonal density matrix elements between states $(\ket{g,1}$, $\ket{g,0})$ and $(\ket{s,1}$, $\ket{s,0})$, respectively. 
Substituting $\expval{\hat{a}}$ in Eq.~(\ref{eq: transmission}), we find
\begin{equation}
    T(\delta)= \abs{\frac{i\delta +\kappa e^{-Dt}}{i\delta +\kappa } -\frac{( i\delta +\Gamma /2) \kappa e^{-Dt}}{g_c^2 +( i\delta +\kappa )( i\delta +\Gamma /2)}}^{2}.
\label{eq: whole transmission}
\end{equation}
The instantaneous transmission resembles that of a cavity QED system [black curve in Fig.~\ref{fig: transmission-one mode}(b)] with an EIT window near $\delta=0$ and two absorption dips at $\delta = \pm g $ separated by the vacuum Rabi frequency; $T(\delta)$ eventually evolves to be $\delta^{2} / (\delta^{2} + \kappa^{2})$, the transmissivity of an empty resonator.

At zero detuning, the transmissivity $T(0)$ decays exponentially with increasing input photon number $\mathcal{I}t$ as
\begin{equation}
    T(0)=T_0e^{-2D_{\text{res}}t} \overset{C \gg 1}{\approx} \exp(-\frac{2 (1-f_{FC})\mathcal{I} t}{C}),
\label{eq: resonant trans}
\end{equation}
which is robust against decay when $C\gg1$. Figure~\ref{fig: transmission-one mode}(a) illustrates sample transmission curves at different cooperativity parameters $C$.

\begin{figure*}[t]
    \centering
    \includegraphics[width=2 \columnwidth]{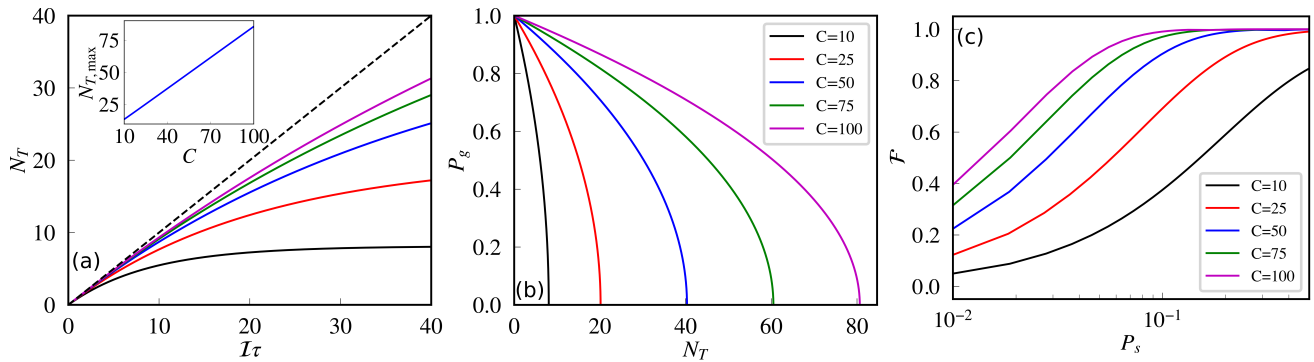}
    \caption{(color online) (a) Transmitted photon number $N_{T}(\tau)$ versus input photon number $\mathcal{I} \tau$ (solid curves from bottom to top) with cooperativity parameters as indicated in the legend of (b). 
    Dashed line represents the upper bound $N_T = \mathcal{I} \tau$. Inset shows $N_{T,\text{max}} \equiv N_T(\infty)$. (b) Ground state molecule population $P_g$ versus $N_{T}$. (c) State detection fidelity $\mathcal{F}$ versus depumping probability $P_s$ for an overall photon counting efficiency $\eta = 0.3$ and a dark-count rate $\mathcal{I}_{dark} = 100$~Hz.}
    \label{fig: photon number}
\end{figure*}

As transmission measurement typically involves finite integration time, we calculate the time-averaged transmission spectra 
\begin{equation}
    \overline{T}(\delta,\tau) = \displaystyle \frac{1}{\tau} \int_{0}^{\tau} T(\delta) dt
\label{eq: time-averaged trans}
\end{equation}
under various time intervals $(0,\tau)$. Figure~\ref{fig: transmission-one mode}(b) shows how the transmitted signal at various laser detuning $\delta$ evolves with the integration time $\tau$. These spectra demonstrate the transition from an EIT-like behavior in a molecule-coupled resonator to the resonant absorption spectrum in an empty resonator -- Two initial transmission dips and a transparency window near $\delta = 0$, formed by the destructive interference of two molecule-photon dressed states, gradually fade away to be overtaken by the single resonance of an empty resonator. Apparently the transmission signal at zero detuning or near the two dips corresponding to the dress-state resonances particularly provides sensitive transient signal for the detection of a molecule in the coupled ground state $\ket{g}$.

We consider background-free transmission at zero detuning since it allows us to take the interrogation time $\tau \rightarrow \infty$ and maximize the number of transmitted photons for detection.
Figure \ref{fig: photon number}(a) displays the relationship between transmitted photon number $N_{T}(\tau)$ and input photon number $\mathcal{I} \tau$, where 
\begin{equation}
    N_{T}(\tau) = \overline{T}(0,\tau) \mathcal{I} \tau \overset{C\gg 1}{\approx} \frac{C}{2 (1-f_{FC})} \left( 1 - e^{-2D_{\text{res}} \tau} \right).
\label{eq: one_trans_photon_number}
\end{equation}
The diagonal dashed line in Fig.~\ref{fig: photon number}(a) represents the optimal case of unity transmissivity for $C \rightarrow\infty$ and for a closed transition $f_{FC} = 1$.
All other cases of finite $C$ and $f_{FC}<1$ fall short of the optimal case and gradually saturates at a maximum photon number
\begin{equation}
N_{T,\text{max}}\equiv N_T(\tau\rightarrow\infty) \overset{C\gg 1}{\approx}\frac{C}{2 (1-f_{FC})},\label{eq: one_max_number}
\end{equation}
indicating that the molecule eventually decouples from the resonator due to pumping to $\ket{s}$. Figure~\ref{fig: photon number}(a) inset shows the approximate linear dependence of $N_{T,\text{max}}$ on the cooperativity $C$ when $C \gg 1$.

On the other hand, for a background-free setup with total single photon counting efficiency $\eta$, the threshold photon number $N_\mathrm{th}$ for molecule detection can be made far less than $N_{T,\text{max}}$. It is possible to conduct a nearly non-destructive state measurement, that is, preserving the initial molecular state following state detection. 

\begin{figure}[t]
    \centering
    \includegraphics[width=0.7 \columnwidth]{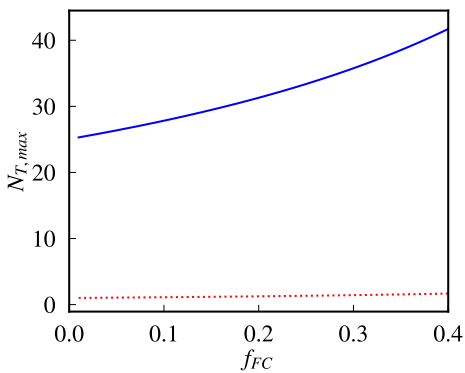}
    \caption{Maximum transmitted photon number $N_{T,max}$ for transitions with different Franck-Condon factors $f_{FC}$ in the cases of cooperativity $C=50$ (solid line). Collectable photon number for fluorescence scattering in freespace is plotted for comparison (dashed line), assuming perfect collection efficiency.}
    \label{fig: f_fc}
\end{figure}

In Fig.~\ref{fig: photon number}(b-c), we calculate the relationship between the transmitted photon number $N_T(\tau)$, the depumping probability $P_s \approx 1-P_g$, and the estimated detection fidelity. We consider approximate Poisson distributions in both the background counts (mean number $\bar{n}_s = \mathcal{I}_\mathrm{dark} \tau$), when a molecule is uncoupled, and the signal photon counts (mean number $\bar{n}_g=\eta N_T(\tau) + \bar{n}_s$), when a molecule is in state $\ket{g}$. Here, $\mathcal{I}_\mathrm{dark}$ is the dark-count rate of a single-photon detector. We define the state detection fidelity for successfully detecting the molecular state as 
\begin{equation}
    \mathcal{F} \equiv \mathrm{min} \left\{p_g(n_\mathrm{th}), p_s(n_\mathrm{th})\right\},
\end{equation}
where $n_\mathrm{th}$ is the threshold photon count that maximizes the fidelity. Here, $p_g(n) = \sum_{k=n}^\infty \bar{n}_g^k e^{-\bar{n}_g}/k!$ and
$p_s(n) = \sum_{k=0}^n \bar{n}_s^k e^{-\bar{n}_s}/k!$. As shown in Fig.~\ref{fig: photon number}(c), with typical experiment parameters ($C=50$ and $\eta=0.3$), near-unity fidelity of 95\% can be achieved with $\sim 15$\% depumping probability and $N_T\approx 10$ as in Fig.~\ref{fig: photon number}(b).

Lastly, we comment that as large cooperativity $C>1$ guarantees higher transmitted photon counts, Eq.~(\ref{eq: one_max_number}) suggests that it is also possible to detect a molecular state using a transition of a small Frank-Condon factor. Figure~\ref{fig: f_fc} shows that the transmitted photon number $N_{T,\text{max}} \sim C/2$ even for $f_{FC} \ll 1$. In comparison, the collectable photon number in direct fluorescence imaging is $1/(1-f_{FC})$ without repumping, subject to finite collection efficiency due to limited solid angle span of imaging instrument.

To sum up, in this section we propose a background-free state detection method, using single WGM coupled to a molecule without a closed transition. An alternative scheme using a single-mode Fabry-Perot cavity has also been investigated. To achieve near background-free detection in a cavity, one should instead monitor the transmissivity at the resonance of a molecule-cavity dressed state. A fiber Fabry-Perot cavity, for example, has a record $C=145$ \cite{colombe2007strong} that would serve as an excellent candidate for the proposed scheme. Details of the adaptation are described in Appendix \ref{app: FB cavity}. 

\begin{figure*}[t]
    \centering
    \includegraphics[width=2 \columnwidth]{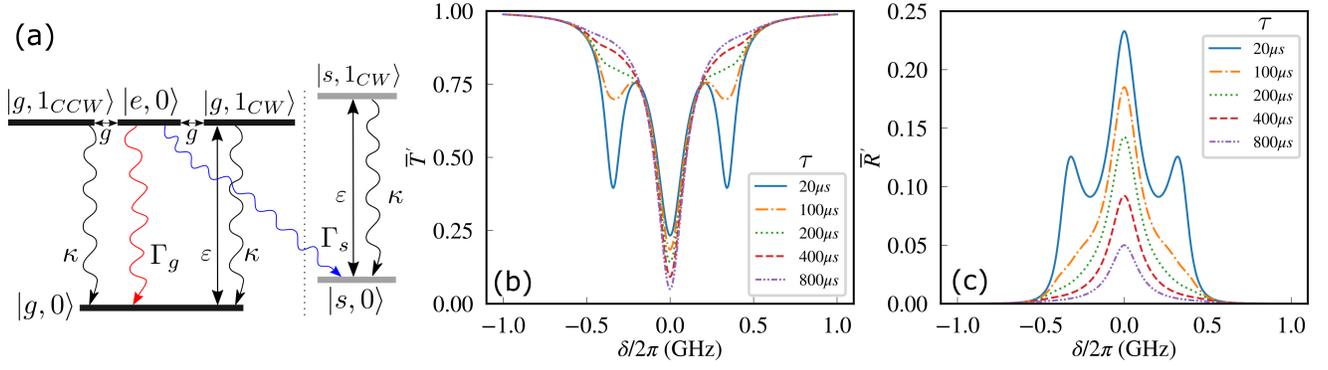}
    \caption{(a)Simplified energy level structure for one molecule coupled to two WGMs in the weak-driving regime. (b) Time-averaged transmission and (c) reflection spectra for probing the molecule under various time intervals with $C=50$.}
    \label{fig: two_weak_driving}
\end{figure*}

\section{\label{sec:two mode}Single Molecule Dynamics Coupled with Two Resoantor Modes} 
In the previous section, we consider single-mode interaction with a spin-polarized molecule. In general, a spin unpolarized molecule can couple to both CW and CCW WGMs. We now consider a general case that a molecule can couple to both modes. We continue to assume negligible back scattering or mode mixing to simplify the discussion. Here, the two modes are degenerate and we model the empty resonator with a two-mode Hamiltonian
\begin{equation}
\begin{aligned}
    \hat{H}^{\prime}_{0} = & \Delta _{cl} (\hat{a}_{CW}^{\dagger } \hat{a}_{CW} + \hat{a}_{CCW}^{\dagger } \hat{a}_{CCW}) \\
    & +i\left(\varepsilon \hat{a}^{\dagger }_{CW} -\varepsilon^{\ast } \hat{a}_{CW}\right)
\end{aligned}
\label{eq: two_H_ring}
\end{equation}
where $\hat{a}_{CW}^{(\dag)}$ and $\hat{a}_{CCW}^{(\dag)}$ are annihilation (creation) operators for the CW and CCW modes, respectively, and we consider the input field in the bus waveguide excites only the CW mode as in Fig.~\ref{fig: ring resonator}(a).

To illustrate the key signature of molecule-WGM-bus waveguide coupling while keeping the calculation tractable, we continue to use a simple two-level structure to effectively model an unpolarized molecule equally coupled to the CW and CCW modes \footnote{While the hyperfine and magnetic sub-level structure should be taken into account in a realistic model, the example given here assumes a simple spherical dipole coupled to the WGMs as an approximation of an unpolarized molecule interacting with the electric field of WGMs.}. We write down the two-mode light-molecule interaction Hamiltonian
\begin{equation}
\begin{aligned}
    \hat{H}^{\prime}_{1} = & \Delta _{ml} \hat{\sigma}_{+} \hat{\sigma} _{-} +i g_{CW}\left( \hat{a}^{\dagger }_{CW} \hat{\sigma} _{-} -\hat{a}_{CW} \hat{\sigma} _{+}\right)\\
     & \ +i g_{CCW}\left( \hat{a}^{\dagger }_{CCW} \hat{\sigma}_{-} -\hat{a}_{CCW} \hat{\sigma}_{+}\right),
\end{aligned}
\label{eq: two_H_mol}
\end{equation}
and assume equal coupling strength with the two modes $g_{CW}=g_{CCW}=g_c$. The master equation of the full system is then modified to be
\begin{equation}
\begin{aligned}
    \frac{d\rho}{dt} = &-i \left[ \hat{H}^{\prime}_{0} + \hat{H}^{\prime}_{1}, \rho \right] + 2\kappa\mathcal{L}[\hat{a}_{CW}]\rho \\ &+2\kappa\mathcal{L}[\hat{a}_{CCW}]\rho  +\Gamma_{g} \mathcal{L}[\hat{\sigma}_{-}]\rho + \Gamma_{s} \mathcal{L} [\hat{\sigma_{-}}^{\prime}]\rho,  
\end{aligned}
\end{equation}
where the two WGMs are assumed to have the same intrinsic loss rates $\kappa_{i}$ and bus waveguide coupling rates $\kappa_{e}$. 

In the limit of single excitation, the resonator and the molecule form an effective six-level system shown in Fig.~\ref{fig: two_weak_driving}(a). State $\ket{g, 1_{CW(CCW)}}$ represents the degenerate level with one photon in the CW (CCW) mode and the molecule in $\ket{g}$. The four coupled states on the left of Fig.~\ref{fig: two_weak_driving}(a) form a cavity QED subsystem with quasi-steady equilibrium, whose population is gradually transferred, via spontaneous decay, to the right part of Fig.~\ref{fig: two_weak_driving}(a) that evolves like an empty resonator described in Sec.~\ref{sec:setup}. The decay rate is similarly described by Eq.~(\ref{eq: decay rate_detuning}) except now $g_c^2$ is replaced by $g^2_{CW}+g^2_{CCW}=2g_c^2$, effectively giving a total cooperativity $2C$. We now have 
\begin{equation}
D^{\prime}(\delta) =\frac{g_c^2 \kappa \mathcal{I} \Gamma_{s}}{\abs{2 g_c^2 +(i \delta +\kappa )\left(i \delta +\frac{\Gamma }{2}\right)} ^{2}},
\label{eq: two_decay_rate}
\end{equation}
based on Eq.~(\ref{eq: two_pop2211}) and assuming $\delta=\Delta_{\text{ml}}=\Delta_{\text{cl}}$ and $\kappa=2 \kappa_e$. The transfer rate at zero detuning is approximately four times slower than that expected in Eq.~(\ref{eq: decay rate}),
\begin{equation}
    D^{\prime}_{\text{res}} =\frac{4C}{( 4C+1)^{2}}( 1-f_{FC})\mathcal{I} \Gamma \overset{C\gg 1}{\approx}\frac{( 1-f_{FC})\mathcal{I}}{4C},
\end{equation}
due to the increased cooperativity $2C$.

While waveguide transmission is modified with the presence of a single molecule, there is now also reflection in the bus waveguide due to the molecule-excited CCW resonator field which couples to the bus waveguide in the backward direction relative to the input field. As similarly discussed in Sec.~\ref{sec:one mode}, we evaluate the time-dependent transmissivity and reflectivity using $T^{\prime} =\left| 1+i\sqrt{\frac{2\kappa _{e}}{\mathcal{I}}} \langle \hat{a}_{CW} \rangle \right| ^{2}$ 
and 
$R^{\prime} =\left| i\sqrt{\frac{2\kappa _{e}}{\mathcal{I}}} \langle \hat{a}_{CCW} \rangle \right| ^{2}$ 
with $\expval{\hat{a}_{CW}}$ and $\expval{\hat{a}_{CCW}}$ calculated in Eqs.~(\ref{eq: exp_a_CW}) and (\ref{eq: exp_a_CCW}) respectively.  
We find
\begin{equation}
    T^{\prime}(\delta) = \abs{1 - \frac{\kappa}{i \delta + \kappa} + r'}^2~\mathrm{and}~R'=\abs{r'}^2,
\label{eq: two_trans}
\end{equation}
where
\begin{equation}
    r'= \frac{ \kappa g_c^2  e^{-D^{\prime} t}}{(i \delta  +\kappa ) \left[ 2 g_c^2 +(i \delta +\kappa )\left(i \delta +\frac{\Gamma }{2}\right) \right]}.
\label{eq: two_refl}
\end{equation}

In Fig.~\ref{fig: two_weak_driving}(b-c), we calculate the time-averaged transmission and reflection spectra at critical coupling, using the definitions similar to Eq.~(\ref{eq: time-averaged trans}) for $\overline{T}^{\prime}(\delta,\tau)$ and $\overline{R}^{\prime}(\delta,\tau)$. There are now three absorption dips (reflection peaks) found in $\overline{T}^{\prime} (\overline{R}^{\prime})$,  resulting from the resonances of three eigenstates within the single excitation subspace of the Hamiltonian $H_0^\prime+H_1^\prime$. Compared to Fig.~\ref{fig: transmission-one mode}(b), 
the molecule-induced transparency and finite reflectivity near $\delta=0$, albeit with much reduced contrast, continue to provide ideal background-free signal. 

\begin{figure}[b]
    \centering
    \includegraphics[width=0.8\columnwidth]{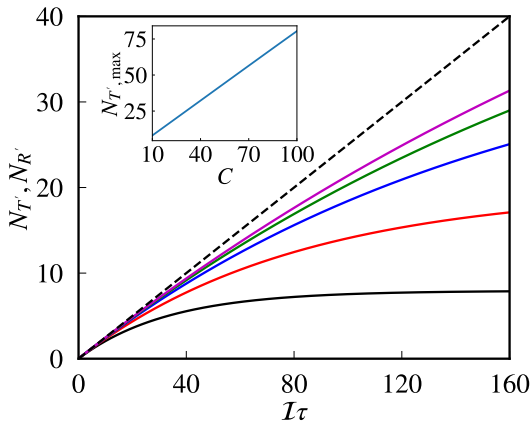}
    \caption{(color online) Transmitted and reflected photon number ($N_{T'}(\tau) = N_{R'}(\tau)$)  versus input photon number $\mathcal{I} \tau$, integrated over time interval $(0,\tau)$, and calculated using $C=1$ (black), 25 (red), 50 (blue), 75 (green), and 100 (purple), respectively (solid curves from bottom to top). Dashed line represents the upper bound $N_{T'} = N_{R'} = \mathcal{I} \tau$. Inset displays $N_{T',\text{max}} \equiv N_{T'}(\infty)$.}
    \label{fig: two_photon number}
\end{figure}

At $\delta=0$, we find equal transmissivity and reflectivity,
\begin{equation}
\begin{aligned}
    T^{\prime}(0)=R^{\prime}(0) &= \abs{\frac{C}{2 C + \frac{1}{2}}}^{2} e^{-2 D^{\prime}_{\text{res}} t} \\
    &\overset{C \gg 1}{\approx} \frac{1}{4} \exp(- \frac{(1- f_{FC}) \mathcal{I} t}{2 C}),
\end{aligned}
\label{eq: two_resonant_tran}
\end{equation}
where the peak values at $t=0$ is reduced to $\approx 25\%$ of the maximum transmission for the single mode case Eq.~(\ref{eq: resonant trans}). The reduced total bus waveguide output, $T^{\prime}(0) + R^{\prime}(0) \lesssim 0.5 < 1$, is due to the excitation of a pure photonic eigenmode that dissipates through the intrinsic resonator loss. Nevertheless, the transfer rate $D^{\prime}_{\textrm{res}}$ in $T^{\prime}(0) \ (  R^{\prime}(0) )$ is also smaller by four times resulting from the collective coupling strength of two modes. As a result, at zero detuning the integrated transmitted (reflected) photon number $N_{T^{\prime}}(\tau)$ ($N_{R^{\prime}}(\tau)$), defined as in Eq.~(\ref{eq: one_trans_photon_number}), still leads to the same maximum counts $N_{T^{\prime},\text{max}} = N_{R^{\prime},\text{max}} = N_{T,\text{max}}$ as shown in  Figs.~\ref{fig: two_photon number} and \ref{fig: photon number}(a).

As suggested from the discussions above, the fidelity for molecular state measurement, using either tranmission or reflection signal alone, remains identical to the case of coupling to a single mode as shown in Fig.~\ref{fig: photon number}(c). 
Moreover, simultaneous detection of bus waveguide transmission and reflection can offer superior sensitivity, with one time more signal photons and with non-trivial temporal correlation between the transmitted and reflected photons when one exploits quantum nonlinearity in the molecule-WGM interactions. 

\begin{figure}[t]
    \centering
    \includegraphics[width=1 \columnwidth]{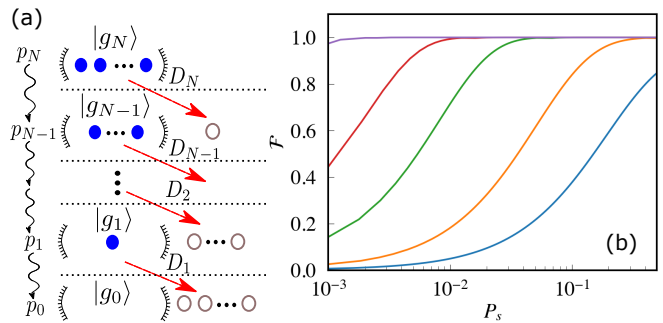}
    \caption{(color online)(a) Effective cascade model describing the population dynamics. When interacting with a weakly-driven resonator, ground state molecules (filled circles) are slowly depumped into the uncoupled states (open circles) one-by-one. $p_{n}$ marks the probability of having $n$ remaining molecules coupled to the resonator and $D_n$ is the transfer rate from $n$ to $n-1$ molecule manifold. (b) State detection fidelity $\mathcal{F}$ versus ground state population loss $P_s$, calculated using $C = 10$ and the initial ground state molecule numbers $N=1$, 2, 5, and 10, 25, respectively (solid curves from bottom to top).}
    \label{fig: cascade model}
\end{figure}

\section{\label{sec:multi}multiple molecule dynamics}
In the previous sections, we assume large cooperativity ($C\gg1$) for single molecule detection. In this section, we discuss the case when more than two molecules in the same state are present in the system. Even with a small cooperativity, one may take advantage of collective effects to achieve state detection with high-fidelity. 

We begin with $N$ ground state molecules trapped on a micro-ring resonator, interacting with a single WGM with identical coupling strength $g_c$. We consider the resonant case $\delta = 0$, and write down the light-molecule interaction Hamiltonian as 
\begin{equation}
\begin{aligned}
\hat{H}_{N} = \sum ^{N}_{\alpha=1} i g_c\left( \hat{a}^{\dagger } \hat{\sigma} ^{\alpha}_{-} e^{-i \phi_{\alpha}} -\hat{a} \hat{\sigma} ^{\alpha}_{+} e^{i \phi_{\alpha}} \right),\label{eq: HN}
\end{aligned}
\end{equation}
where the index $\alpha$ labels individual molecules and $\phi_\alpha$ represents 
a position-dependent phase in the light-molecule coupling since a WGM is a traveling wave.

We incorporate a modified Dicke model to investigate the collective behavior of these resonator-coupled molecules. Let 
$\displaystyle \hat{J}^{N}_{\pm} = \frac{1}{\sqrt{N}}\sum ^{N}_{\alpha=1} \hat{\sigma} ^{\alpha}_{\pm} e^{\pm i \phi_{\alpha}}$
be the collective spin lowering and raising operators, we can rewrite the equivalent Hamiltonian for Eq.~(\ref{eq: HN}) as 
\begin{equation}
    \hat{H}_{J_N} =  i \sqrt{N} g_c \left( \hat{a}^{\dagger } \hat{J}^{N}_{-} -\hat{a} \hat{J}^{N}_{+}\right),
\end{equation}
where the molecule-resonator coupling strength is replaced by $\sqrt{N}g_c$.

In the limit of single excitation, the resonator-coupled $N$ molecules resemble an effective two-level system with a $N$-molecule ground state $\ket{g_N} \equiv \Pi^N_{\alpha=1}  \ket{g}_\alpha$ and an excited state $\ket{e_N} \equiv \hat{J}^{N}_{+}\ket{g_N}$, in which one molecule gets excited to $\ket{e}$. Spontaneous decay (via emitting single photon into freespace) can either bring the population in $\ket{e_N}$ back to $\ket{g_N}$ or bring one excited state molecule down to the uncoupled state $\ket{s}$. The full master equation for the resonator-coupled $N$-molecule density matrix $\rho_N$ can be expressed as
\begin{equation}
\begin{aligned}
    \frac{d\rho _{N}}{dt} = & -i[\hat{H}_{0} + \hat{H}_{J_N} ,\rho _{N}] + \Gamma _{g}\mathcal{L}\left[\hat{J}^{N}_{-}\right] \rho_{N} \\
 & + 2\kappa \mathcal{L}[\hat{a}] \rho_{N}  + \Gamma _{s}\sum_{\alpha=1}^N\mathcal{L}\left[\hat{\sigma}^{\prime \alpha}_{-} e^{-i \phi_{\alpha}} \right] \rho _{N},\label{eq: NmoleculeMaster}
\end{aligned}
\end{equation}
where the first line of the equation describes the evolution of the $N$-molecule collective states with coupling strength $\sqrt{N}g_c$ and decay rate $\Gamma_g$, and the second line adds resonator dissipation and single molecular decay into $\ket{s}$.

While full evolution dynamics of Eq.~(\ref{eq: NmoleculeMaster}) can be evaluated numerically, calculation for large $N$ can be computationally expansive. Here, we develop an analytical approximation in the weak-driving limit. Starting with $N$ molecules in the ground state $\ket{g_N}$ weakly excited to $\ket{e_N}$ by the resonator mode, the system evolves collectively similar to the single molecule case in Fig.~\ref{fig: one_weak driving}(a). Within this $N$-coupled molecule manifold, the system can be described by a simple three level system consisting of  $\ket{g_N, 0}$, $\ket{g_N, 1}$ and $\ket{e_N, 0}$ until spontaneous decay into state $\ket{s}$ occurs. If we trace out the molecule that decays into the uncoupled state, not knowing which one did, the system can be described again by a collective state $\ket{g_{N-1}}$ with $N-1$ molecules in the ground state, weakly excited to $\ket{e_{N-1}}$, as detailed in Appendix \ref{app: multi}. The dynamics can cascade down as prescribed with $N-2, N-3, ...,1$ molecule(s) left in the system until all molecules become uncoupled, as illustrated in Fig.~\ref{fig: cascade model} (a).

We can calculate the probability $p_n$ of having $n$ coupled molecules in the system, using the simple cascade model. This leads to a system of equations,
\begin{equation}
\begin{cases}
\displaystyle \frac{dp _{N}}{dt} =-p _{N} D_{N}\\
\displaystyle \frac{dp _{n}}{dt} =p _{n+1} D_{n+1} -p _{n} D_{n},  \ 1< n <N\\
\displaystyle \frac{dp _{0}}{dt} =p _{1} D_{1},
\end{cases}
\label{eq: cascade model}
\end{equation}
where $p_0$ is for all molecules in $\ket{s}$. The effective transfer rate of the population from $n$ to $n-1$ coupled molecules can be calculated according to Eq.~(\ref{eq: p_en0gn0}).
We find 
\begin{equation}
D_{n} =  \frac{2 n g_c^2 \kappa_{e} \mathcal{I}}{\abs{n g_c^2 +\kappa \frac{\Gamma }{2}} ^{2}}\Gamma_{s} \label{eq: N decay rate_detuning}
\end{equation}
as detailed in Appendix~\ref{app: multi}. For $n \geq 1$, the transfer rate is suppressed by the cooperativity $\sim 1/nC$. We note that Eq.~(\ref{eq: decay rate_detuning}) (with $\delta=0$) is the single molecule case of Eq.~(\ref{eq: N decay rate_detuning}). 

We can also derive the bus waveguide transmission by finding the expectation value for the resonator field,
\begin{equation}
    \displaystyle \expval{\hat{a}} =\sum_{n=0}^{N}\frac{\frac{\Gamma }{2}}{n g_c^2 +\kappa \frac{\Gamma }{2}}  p_{n} \varepsilon.
    \label{eq: multiN mode}
\end{equation}
Bus waveguide transmissivity $T_N$ for initially $N$ ground state molecules can then be evaluated using Eq.~(\ref{eq: transmission}).

We have compared the analytical solutions with numerical calculations for mean populations and photon numbers, and found very good agreement for $N=2,3,4$. Equations~(\ref{eq: cascade model}-\ref{eq: multiN mode}) allow us to evaluate the dynamics of waveguide transmission with arbitrarily large number of coupled molecules.

The major advantage for coupling more than one molecules to a resonator is that the collective coupling leads to many more signal photons and higher fidelity without losing significant fraction of ground state molecules to the uncoupled states. To illustrate this, in Fig.~\ref{fig: cascade model}(b) we calculate the state detection fidelity $\mathcal{F}$ as a function of the ground state molecule loss
\begin{equation}
    P_s = 1 - \frac{1}{N}\sum_{n=1}^{N} n p_n.
\end{equation}
It is shown that, under a moderate $C = 10$, $N=10$ ground state molecules can be detected with over 99\% fidelity with 1\% ground state population loss. 
For even larger $N$, non-destructive state detection with negligible loss can be realized with cooperavity parameter $C<0.1$, as described in Ref.~\cite{sawant2018detection}.

\section{Conclusion}
To conclude, we have proposed a background-free state detection scheme for single molecules without an optically closed transition. High-fidelity measurement can be realized in resonators with a sufficiently large cooperativity $C > 10$. A possible experiment with cold molecules could begin with an array of cold atoms trapped in optical tweezers \cite{thompson2013coupling, kim2019trapping} or in a lattice of evanescent field traps above the surface of a high-Q micro-ring resonator or a photonic crystal cavity of $C \gtrsim 25$ \cite{chang2019microring, Samutpraphoot2020Strong}. Resonator-assisted photoassociation (PA) to a molecular ground state (also with high fidelity) can be performed  by simply introducing PA light into the experimental setup \cite{perez2017ultracold, kampschulte2018cavity}. Immediately following PA, one can detect the existence of ground state molecules using the proposed scheme with probe photons directly launched into the bus waveguide. This state detection technique could also be employed for atoms \cite{kim2019trapping} or quantum emitters coupled to a high-Q micro-ring resonator or other whispering-gallery mode resonators.

\begin{acknowledgments}
We thank J. P\'{e}rez-R\'{\i}os for discussions. Funding is provided by the Office of Naval Research (N00014-17-1-2289). M. Zhu acknowledges support from the Rolf Scharenberg Graduate Fellowship. C.-L. Hung acknowledges support from the AFOSR YIP (FA9550-17-1-0298).
\end{acknowledgments}

\appendix
\renewcommand\thefigure{\thesection\arabic{figure}} 
\setcounter{figure}{0} 
\section{\label{app: FB cavity} Alternative setup using a Fabry-Perot cavity} 
\begin{figure}[b]
    \centering
    \includegraphics[width=1 \columnwidth]{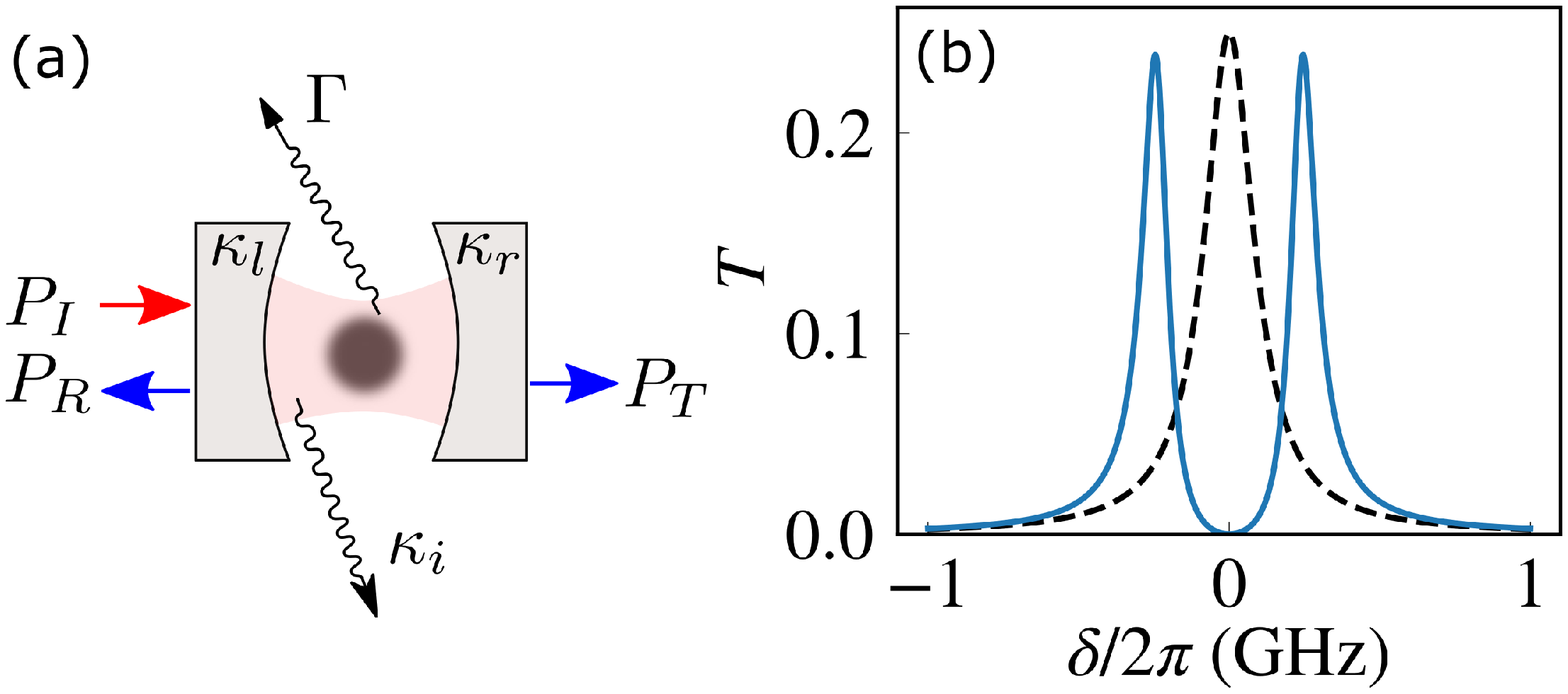}
    \caption{(a) Schematic illustration of the coupling between a Fabry-Perot cavity and a molecule. (b) Transmission spectra with (solid line) and without a ground state molecule coupled to the cavity(dashed line).}
    \label{fig: FB cavity}
\end{figure}

Fabry-Perot cavity is widely used in the investigation of cavity QED. In this section we discuss a similar way to detect a molecule with no optically closed transitions. The system Hamiltonian is similarly described by Eq.~(\ref{eq: H_ring}), except now the driving field amplitude $\varepsilon = \sqrt{2 \kappa_{l} \mathcal{I}}$, where $\kappa_{l(r)}$ is the effective loss rate due to transmission through the left (right) mirror, as illustrated in Fig.~\ref{fig: FB cavity}(a) and $\kappa=\kappa_{i}+\kappa_{r}+\kappa_{l}$ is the cavity total decay rate.

Figure \ref{fig: FB cavity}(b) displays the transmission spectra with and without a molecule, assuming a short interrogation time, and with the parameters listed in Table~\ref{tab: parameters} where $\kappa_r = \kappa_l = 2\pi \times 25\text{MHz}$. In the cavity setup, transmissivity is
$\displaystyle T=\abs{\sqrt{\frac{2 \kappa_{r}}{\mathcal{I}}} \expval{\hat{a}}}^{2}$
, where the time evolution of $\expval{\hat{a}}$ can be similarly evaluated as in Sec.~\ref{sec:one mode}. When a molecule in the target state is coupled to the cavity, transmission around the resonance splits into two peaks due to the vacuum Rabi splitting. One could thus monitor the cavity transmission at $\delta =\pm g_c$, where the transmissivity increases by more than ten-fold, to perform nearly background-free measurement. 

\section{\label{app: single} Derivation for the quasi-steady state density matrix in the case of one resonator mode}
In the weak-driving regime, we derive the effective decay rate from the cavity QED subsystem to the empty resonator states as shown in Fig.~\ref{fig: single dynamics}(a). Here, we determine the elements of the quasi-steady density matrix $\rho^{ss}$. We introduce an arbitrarily slow artificial repump of rate $\zeta$ between $\ket{s,0}$ ($\ket{s,1}$) and $\ket{g,0}$ \cite{kampschulte2018cavity}. For convenience, we define $\ket{1} \equiv \ket{g,0}$, $\ket{2} \equiv \ket{e, 0}$, $\ket{3} \equiv \ket{g,1}$, $\ket{4} \equiv \ket{s,0}$,  $\ket{5} \equiv \ket{s,1}$, and $\hat{\sigma}_{ij} \equiv \ket{i}\bra{j}$. Then the Hamiltonian of the full system with artificial repump is

\begin{equation}
    \hat{H}_{R} = \hat{H}_{0} + \hat{H}_{1} + \zeta \left(\hat{\sigma}_{41} + \hat{\sigma}_{14} + \hat{\sigma}_{51} + \hat{\sigma}_{15} \right)
\end{equation}
where $\hat{H}_{0}$ is the Hamiltonian of empty resoantor as in Eq.~(\ref{eq: H_ring}) and $\hat{H}_{1}$ is the single-mode light-molecule interaction Hamiltonian as in Eq.~(\ref{eq: H1}). Taking into account the loss channels, we write down the master equation
\begin{equation}\label{eq:master_one}
    \frac{d\rho}{dt} = -i \left[ \hat{H}_{R}, \rho \right] + 2\kappa\mathcal{L}[\hat{a}]\rho +\Gamma_{g} \mathcal{L}[\hat{\sigma_{12}}]\rho + \Gamma_{s} \mathcal{L}[\hat{\sigma_{42}}]\rho.
\end{equation}

We first focus on the evolution of the density matrix elements to the leading order of $\abs{\varepsilon}$ (assuming $\zeta \ll \varepsilon $). We define $\rho_{ij}=\bra{i}\hat{\rho}\ket{j}$, and find the evolutions of for $\rho_{21}$ and $\rho_{31}$ satisfy 
\begin{equation}
\begin{aligned}
    \frac{d\rho _{21}}{d t} &=-\ ( i \Delta_{\text{ml}} +  \frac{\Gamma}{2}  )\rho _{21}  -g_c\rho _{31},\\
    \frac{d \rho_{31}}{d t} &= -\ (i  \Delta_{\text{cl}}+\kappa) \rho_{31}  + \varepsilon  \rho_{11}  + g_c \rho_{21},
\end{aligned}
\label{single-dynamics}    
\end{equation}
which is independent of $\zeta$ as well as the empty resonator states $\ket{4}$ and $\ket{5}$. Solving for the quasi-steady state, $\displaystyle \frac{d \rho_{31}}{d t} \approx \frac{d\rho _{21}}{d t} \approx 0$, we find 
\begin{align}
    \frac{\rho _{21}^{ss}}{\rho _{11}^{ss}} &=\frac{-g_c}{g_c^2 +(i \Delta _{\text{cl}} +\kappa )\left(i  \Delta_{\text{ml}} +\frac{\Gamma }{2}\right)} \varepsilon \\
    \frac{\rho _{31}^{ss}}{\rho _{11}^{ss}} &=\frac{i  \Delta_{\text{ml}} +\frac{\Gamma }{2}}{g_c^2 +(i \Delta _{\text{cl}} +\kappa )\left(i  \Delta_{\text{ml}} +\frac{\Gamma }{2}\right)} \varepsilon. \label{eq: rho31_single}
\end{align}
We similarly derive the components of $\rho^{ss}$ to the next order of $\abs{\varepsilon}$, 
\begin{equation}
\frac{\rho _{22}^{ss}}{\rho_{11}^{ss}} =\frac{2 g_c^2  \kappa_e \mathcal{I}}{\abs{ g_c^2 +(i \Delta _{\text{cl}} +\kappa )\left(i  \Delta _{\text{ml}} +\frac{\Gamma }{2}\right)} ^{2}} 
\label{eq: e0_pop}
\end{equation}
\begin{equation}
    \frac{\rho _{33}^{ss}}{\rho_{11}^{ss}} =\frac{2 \left((\Gamma /2)^{2} + \Delta_{\text{ml}}^{2}\right) \kappa_e \mathcal{I}}{\abs{ g_c^2 +(i \Delta _{\text{cl}} +\kappa )\left(i  \Delta _{\text{ml}} +\frac{\Gamma }{2}\right)} ^{2}}.
\end{equation}
where we have used $\varepsilon = i \sqrt{2 \kappa_e \mathcal{I}}$.

Under weak-driving, the initial population in the ground state $\ket{g}$ is gradually transferred to states $\ket{s}$ via the spontaneous decay channel $\ket{2} \rightarrow \ket{4}$. From Eq.~(\ref{eq: e0_pop}), we obtain the population ratio between states $\ket{e,0}$ and $\ket{g,0}$ and find the effective transfer rate $D$ 
\begin{equation}
D =  \frac{\rho_{22}^{ss}}{\rho_{11}^{ss}} \Gamma_{s} = \frac{ 2 g_c^2 \kappa_e \mathcal{I}}{\abs{ g_c^2 +(i \Delta _{\text{cl}} +\kappa )\left(i \Delta _{\text{ml}} +\frac{\Gamma }{2}\right)} ^{2}}\Gamma_{s}, 
\end{equation}
and  Eq.~(\ref{eq: decay rate_detuning}) for the case at critical coupling and $\Delta _{\text{cl}} = \Delta _{\text{ml}}$. Considering initially the system begins in $\ket{g,0}$ and most of the population resides in either state $\ket{1}$ ($\ket{g,0}$) or $\ket{4}$ ($\ket{s,0}$), we find $\rho_{11}(t) \approx e^{-D t}$, $\rho_{44}(t) \approx 1 - e^{-D t}$, and hence Eq.~(\ref{eq: population}). The dynamics of all other components in $\rho$ can be solved once the populations of $\ket{1}$ and $\ket{4}$ are known. \par

We can solve the expectation value $\expval{a}$ for calculating transmissivity in Eq.~(\ref{eq: transmission}).
We find
\begin{equation}
\begin{aligned}
    \expval{\hat{a}} =& \rho_{31} + \rho_{54} \\  
    =& \frac{i  \Delta_{\text{ml}} +\frac{\Gamma }{2}}{g_c^2 +(i \Delta _{\text{cl}} +\kappa )\left(i  \Delta_{\text{ml}} +\frac{\Gamma }{2}\right)} \varepsilon e^{-D t}\\
    &+ \frac{\varepsilon}{i \Delta _{\text{cl}} +\kappa } (1- e^{-D t}).
\end{aligned}
\label{eq: exp_a_one_mode}
\end{equation}
We can also calculate the dynamics of resonator photon number $\expval{\hat{a}^{\dag} \hat{a}}$, using the populations of one photon states
\begin{equation}
\begin{aligned}
    \expval{\hat{a}^{\dag} \hat{a}} =& \rho_{33} + \rho_{55} \\
    =& \frac{2 \left((\Gamma /2)^{2} + \Delta_{\text{ml}}^{2}\right) \kappa_e \mathcal{I}}{\abs{ g_c^2 +(i \Delta _{\text{cl}} +\kappa )\left(i  \Delta _{\text{ml}} +\frac{\Gamma }{2}\right)} ^{2}} e^{-D t}\\
    &+ \frac{2 \kappa_e \mathcal{I}}{\Delta _{\text{cl}}^2 +\kappa^2 } (1- e^{-D t}).
\end{aligned}
\label{eq: photon_number_one_mode}
\end{equation}
Figure~\ref{fig: single dynamics} validates the analytical approximation with full numerical calculations in the weak driving regime.

\section{\label{app: two} Derivation for the quasi-steady state density matrix in the case of two resonator modes}
The same scenario used in Appendix~\ref{app: single} can be applied in the case of coupling to two resonator modes. We can evaluate the transfer rate to the empty resonator state $D^{\prime}$ and the expectation value of $\expval{a_{CW}}$ and $\expval{a_{CCW}}$ in the weak-driving regime. Defining $\ket{1} \equiv \ket{g,0}$, $\ket{2} \equiv \ket{e, 0}$, $\ket{3} \equiv \ket{g,1_{CW}}$, $\ket{4} \equiv \ket{g, 1_{CCW}}$, $\ket{5} \equiv \ket{s,0}$ and $\ket{6} = \ket{s,1}$, we obtain
\begin{equation}
\frac{\rho _{22}^{ss}}{\rho_{11}^{ss}} =\frac{2 g_c^2 \kappa_e \mathcal{I}}{\abs{2 g_c^2 +(i \Delta _{\text{cl}} +\kappa )\left(i \Delta_{\text{ml}} +\frac{\Gamma}{2}\right)} ^{2}} 
\label{eq: two_pop2211}
\end{equation}
and thus the transfer rate $D^{\prime}$ in Eq.~(\ref{eq: two_decay_rate}). \par

Similarly, we find $\rho_{11}(t) = e^{-D^{\prime} t}$, $\rho_{44}(t)=1-e^{-D^{\prime} t}$, 
\begin{equation}
\begin{aligned}
   \expval{\hat{a}_{CW}} =&  \rho _{31} + \rho_{65}  \\  
    =& \frac{g_c^2 +(i \Delta _{\text{cl}} +\kappa )\left(i  \Delta _{\text{ml}} +\frac{\Gamma }{2}\right)}{(i \Delta _{\text{cl}} +\kappa ) \left[ 2 g_c^2 +(i \Delta _{\text{cl}} +\kappa )\left(i  \Delta _{\text{ml}} +\frac{\Gamma }{2}\right) \right]} \varepsilon e^{-D^{\prime} t}\\
    &+ \frac{\varepsilon}{i \Delta _{\text{cl}} +\kappa } (1- e^{-D^{\prime} t}),
\end{aligned}
\label{eq: exp_a_CW}
\end{equation}
\begin{equation}
\begin{aligned}
    \expval{\hat{a}_{CCW}} &= \rho_{41} \\
    &= \frac{-g_c^2}{(i \Delta _{\text{cl}} +\kappa ) \left[ 2 g_c^2 +(i \Delta _{\text{cl}} +\kappa )\left(i  \Delta_{\text{ml}} +\frac{\Gamma }{2}\right) \right]} \varepsilon e^{-D^{\prime} t}.
\end{aligned}
\label{eq: exp_a_CCW}
\end{equation}
Substituting $\expval{\hat{a}_{CW}}$ and $\expval{\hat{a}_{CCW}}$ in the expressions of $T^{\prime}$ and $R^{\prime}$ leads to Eqs.~(\ref{eq: two_trans}) and (\ref{eq: two_refl}).\par

\section{\label{app: multi} Cascade model for multiple molecules resonantly coupled to one resonator mode}
In the main text, we consider multiple molecules collectively couple to one resonator mode by directly tracing out all the position dependence. Here, we explicitly carry out the derivation and arrive at the cascade model described by Eq.~(\ref{eq: cascade model}).

We assume $N$ trapped molecules randomly spread along the micro-ring resonator. We denote $\ket{g_{n,k}}$ as the $k$-th configuration that satisfies  $N-n$  molecules in state $\ket{s}$ and $n$ molecules in state $\ket{g}$, and we define $G_k$ to be the set of positions labeling these $n$ molecules. We assume every molecule in the ground state can be equally excited by the WGM. Thus, with single excitation created in a given configuration $\ket{g_{n,k}}$, the system forms a superposition state $\ket{e_{n,k}, 0} \equiv \frac{1}{\sqrt{n}} \sum_{\alpha \in G_k} \hat{\sigma}^\alpha_+ e^{i \phi_{\alpha}} \ket{g_{n,k}}$. Each excited state molecule might decay to $\ket{s}$, thus $\ket{e_{n,k}, 0}$ can evolve into $n$ orthogonal configurations of $n-1$ molecules in the ground state, $\ket{g_{n-1,k'},0}=\ket{s_\alpha}\bra{g_\alpha}\ket{g_{n,k}, 0}$, and the superposition is destroyed. Here, $\alpha \in G_k$ labels the position of the molecule that decays into $\ket{s}$, and we denote $\ket{g_{n-1,k'}}$ as the $k'$-th configuration that $n-1$ molecules are in state $\ket{g}$. 

We note that the evolution dynamics of each position configuration is identical, since the major differences between the states $\ket{e_{n,k}, 0}$ are the position-dependent phases that can be absorbed as a part of the spin lowering and raising operators defined in Sec.~\ref{sec:multi}. If we ignore the position information, one can trace out all different configurations without losing relevant physical information. For simplicity, we denote the states in analogy to the states in Appendix~\ref{app: single},
\begin{equation}
    \begin{aligned}
    \ket{1_{n,k}} &\equiv \ket{g_{n,k}, 0}\\
    \ket{2_{n,k}} &\equiv \ket{e_{n,k}, 0}\\ 
    \ket{3_{n,k}} &\equiv\ket{g_{n,k}, 1}\\
    \ket{4_{n,k'}} &\equiv\ket{1_{n-1,k'}}
    \end{aligned}
    \label{def123}
\end{equation}
where $k$ and $k'$ are the indices of configurations for $n$ and $n-1$ molecules coupled to the resonator, respectively. We trace out all the $k$ configurations in the density matrix elements by the following summation
\begin{equation}
    \begin{aligned}
    \rho_{i_n, j_n}  &\equiv \sum_k \rho_{i_{n,k}, j_{n,k}},
    \end{aligned}
\end{equation}
for $i,j=1,2,3$ and $\rho_{i_{n,k}, j_{n,k}} \equiv \bra{i_{n,k}} \hat{\rho} \ket{j_{n,k}}$. When states $\ket{4_{n,k'}}$ are involved, we additionally trace out all $k'$ configurations
\begin{equation}
    \begin{aligned}
    \rho_{i_n, 4_n}  &\equiv \sum_{k,k'} \rho_{i_{n,k}, 4_{n,k'}},
    \end{aligned}
\end{equation}
and similar definitions for $\rho_{4_n,i_n}$ and $\rho_{4_n,4_n}$ follow. We note that $\rho_{4_n,4_n}=\rho_{1_{n-1},1_{n-1}}$ for $n\geq1$.

These density matrix elements evolve similar to $\rho_{ij}$ of one molecule coupled to one WGM. For example, we find
\begin{equation}
\begin{aligned}
    \frac{d\rho _{2_n, 1_n}}{d t} &=-\ ( i \Delta_{\text{ml}} +  \frac{\Gamma}{2}  )\rho _{2_n,1_n}  -\sqrt{n}g_{c}  \rho _{3_n, 1_n},\\
    \frac{d \rho_{3_n, 1_n}}{d t} &= -\ (i  \Delta_{\text{cl}}+\kappa) \rho_{3_n, 1_n}  + \varepsilon  \rho_{1_n, 1_n}  + \sqrt{n}g_{c} \rho_{2_n, 1_n},
\end{aligned}
\label{n-dynamics}
\end{equation}
sharing the same form as those in Eq.~(\ref{single-dynamics}) 
except that we replace $g_c$ by $\sqrt{n}g_{c} $. Following the procedures in Appendix \ref{app: single}, we obtain the ratio of quasi-steady state population at zero detuning
\begin{align}
\label{eq: p_gn1gn0}
    \frac{\rho_{g_{n}1,g_{n}0}}{\rho_{g_{n}0,g_{n}0}} \equiv \frac{\rho_{3_n,1_n}}{\rho_{1_n,1_n}} &=\frac{\frac{\Gamma }{2} \varepsilon}{n g_c^2 +\kappa \frac{\Gamma }{2}}\\
    \frac{\rho_{e_{n}0,e_{n}0}}{\rho_{g_{n}0,g_{n}0}}  \equiv \frac{\rho_{2_n,1_n}}{\rho_{1_n,1_n}}&= \frac{2 n g_c^2 \kappa_{e} \mathcal{I}}{\abs{n g_c^2 +\kappa \frac{\Gamma }{2}} ^{2}},
\label{eq: p_en0gn0}
\end{align}
where we have converted the subscripts of the density matrix elements to those used in the main text. Given the ground state population $\rho_{g_{n}0,g_{n}0} \equiv \rho_{1_n,1_n}$, the above equations describe the dynamics within each $n$-coupled molecule manifold.  

Since $\rho_{4_n,4_n}=\rho_{1_{n-1},1_{n-1}}$, spontaneous decay to state $\ket{s}$ connects the population in every $n$-coupled molecule manifold with the populations in the $n\pm1$ manifolds. We thus arrive at the cascade decay model Eq.~(\ref{eq: cascade model}). Solving for the population $\rho_{g_{n}0,g_{n}0}$ in each $n$-molecule manifold, we can also obtain the dynamics of transmission using the expectation value of the resonator mode field, Eq.~(\ref{eq: p_gn1gn0}),
\begin{equation}
    \expval{\hat{a}} = \sum_{n=0}^{N}\rho_{g_{n}1,g_{n}0}.
\label{eq: a_exp_n}
\end{equation}

\bibliography{main}

\end{document}